\documentclass{article}

\PassOptionsToPackage{numbers, compress}{natbib}

    \usepackage[final]{neurips_2021}

\usepackage[utf8]{inputenc} 
\usepackage[T1]{fontenc}    
\usepackage{hyperref}       
\usepackage{url}            
\usepackage{booktabs}       
\usepackage{amsfonts}       
\usepackage{nicefrac}       
\usepackage{microtype}      
\usepackage{xcolor}         
\usepackage{graphicx}
\usepackage{subcaption}
\usepackage{titlesec}

\title{ECG for high-throughput screening of multiple diseases: Proof-of-concept using multi-diagnosis deep learning from population-based datasets}

\author{Weijie Sun$^{1,2}$ \And Sunil Vasu Kalmady$^{1,3}$ \And Amir Salimi$^2$ \And Nariman Sepehrvand$^1$ \And Eric Ly$^1$ \And Abram Hindle$^2$ \And Russell Greiner$^{2,3}$ \And Padma Kaul$^1$ 
\And 
\\[-5pt]
$^1$ Canadian VIGOUR Centre, Department of Medicine, University of Alberta, Alberta, Canada \\
$^2$ Department of Computing Science, University of Alberta, Alberta, Canada \\
$^3$ Alberta Machine Intelligence Institute, Alberta, Canada \\
\AND
weijie2@ualberta.ca \And kalmady@ualberta.ca 
}

\begin{document}

\maketitle

\begin{abstract}

Electrocardiogram (ECG) abnormalities are linked to cardiovascular diseases, but may also occur in other non-cardiovascular conditions such as mental, neurological, metabolic and infectious conditions. However, most of the recent success of deep learning (DL) based diagnostic predictions in selected patient cohorts have been limited to a small set of cardiac diseases. In this study, we use a population-based dataset of >250,000 patients with >1000 medical conditions and >2 million ECGs to identify a wide range of diseases that could be accurately diagnosed from the patient’s first in-hospital ECG. Our DL models uncovered 128 diseases and 68 disease categories with strong discriminative performance.

\end{abstract}

\section{Introduction}

Electrocardiogram (ECG) captures the propagation of the electrical signal in the heart and is one of the most routinely used non-invasive modalities in healthcare to diagnose cardiovascular diseases \cite{lyon2018computational}. However, ECG signals can be complex, making it challenging and time-consuming to interpret, even for experts. In recent years, deep learning (DL) models have been successful in reaching near human levels of performance, however most of these studies have been limited to typical ECG abnormalities such as arrhythmias \cite{alday2020classification} and a limited set of heart diseases including valvulopathy, cardiomyopathy, and ischaemia \cite{somani2021deep}.

Several clinical studies have shown strong associations of ECG abnormalities with numerous diseases beyond cardiovascular conditions, including but not limited to mental disorders : depression \cite{wang2013altered}, bipolar disorder \cite{hage2019low}; infectious conditions : HIV \cite{soliman2011prevalence}, sepsis \cite{shashikumar2017early}; metabolic diseases : diabetes type 2 \cite{harms2021prevalence}, amyloidosis \cite{cheng2013findings}; drug use : psychotropics \cite{polcwiartek2020electrocardiogram}, cannabis \cite{yahud2020cannabis}; neurological disorders: Alzheimer disease \cite{zulli2005qt}, cerebral palsy \cite{pastore2011characterization}; respiratory diseases : pneumoconiosis \cite{wu2019effects}, chronic obstructive pulmonary disease \cite{larssen2017mechanisms}; digestive system diseases : liver cirrhosis \cite{toma2020electrocardiographic}, alcoholic liver disease \cite{wehr1990cardiac}; miscellaneous conditions: chronic kidney disease \cite{shafi2017ecg}, preterm labour \cite{gemelli1990transient}, systemic lupus erythematosus \cite{myung2017prevalence} etc. However, despite well established clinical associations of ECG changes with multiple diseases, very few studies have explored the information contained in ECGs that could be harnessed for prediction of non-cardiovascular conditions. A major challenge here is the lack of availability of large training datasets of digitized ECGs that could be linked to concurrent diagnostic information across various disease types. In this context, standardized administrative health data, routinely generated at each encounter, provide a wonderful opportunity to explore the full spectrum of patient diagnoses. These data include the most responsible diagnosis, as well as any comorbidities the patient may have or develop during presentation.

In this study, we use a population-based dataset of >250,000 patients with various medical conditions and >2 million in-hospital ECGs. Here, we use diagnoses coded using the World Health Organization International Classification of Diseases (ICD) \cite{who2016ICD10}. The goal of our study is to identify which diseases (with previously known or unknown associations with ECGs) can be accurately diagnosed from the patient’s first ECG during an emergency department (ED) visit or hospitalization based on a learned DL model. It aims to provide a proof-of-concept for high-throughput screening of ICD-wide range of diseases based on ECG, and presents disease candidates to be explored in future ECG studies with focused investigation on specific diagnosis.

\section{Method}
This study used population-based datasets from 26 hospitals in Alberta, Canada (2007-2020), containing information on 772,932 healthcare episodes (hospitalization and ED visits) of 260,065 patients who collectively had 13,179 unique ICD-10 codes/diseases \cite{who2016ICD10}. We linked these episodes to a dataset of 2,015,808 ECGs (Philips IntelliSpace system, 12-lead, 500 Hz, 10 s) using unique patient identifiers and timing of ECG acquisition. After data cleaning and exclusions (poor signal quality\footnote{Trace quality was ensured on muscle artifact, AC noise, baseline wander, QRS clipping, leads-off flags etc.}, unlinked episodes, pacemaker and devices, < 18 years old, etc.), we used 1,514,968 ECGs that were linked to 724,074 episodes of 239,852 patients with 11,207 unique ICD codes. An ICD-10 code is 3 to 7 characters that specifies a specific disease, where the first 3 characters denote the general category of disease (e.g., ‘I214’ refers to ‘Non-ST elevation (NSTEMI) myocardial infarction’ and  ‘I21’ refers to its broader category ‘Acute myocardial infarction’). We used ICD codes and corresponding categories as labels for prediction modelling. We found 1,319 ICD codes (full code, exact match) and 699 ICD categories (match first 3 digits) that were each linked to at least 1000 ECGs. 

We split our ECG dataset into the internal validation set (random 60\%: 143,939 patients with 436,508 ECGs, used for training and internal validation) and external holdout set (remaining 40\%: 95,913 patients with 287,566 ECGs), while ensuring that ECGs from the same patient were not shared between the sets. Whenever there were multiple ECGs in an episode, we used only the first ECG for evaluation, as it would be preferable in actual clinical practice to make a diagnostic prediction at the first point of care in the ED or hospital. We trained two DL models, for full ICD codes and ICD categories. We first trained and evaluated the performance with 80\%-20\% split within the internal set, and selected a list of top labels based on discriminative performance (Area under receiver operating characteristic curve (AUROC)). We then retrained the models on the entire internal set and evaluated on the external set based on the selected labels. Our DL architecture was based on ResNet \cite{he2016deep}, similar to the one used in earlier ECG modeling study \cite{ribeiro2020automatic}. Here, 12-lead ECG traces were input to the network, consisting of convolutional layer (conv), 4 residual blocks with 2 conv per block, followed by a dense layer to which age and sex features were concatenated. We used batch normalization, ReLU and dropout after each conv. The last block is then fed into a dense layer with sigmoid activation to output a 1319 (resp., 699) length vector of predicted probabilities for the codes/diseases (resp., categories). We used the Adam optimizer, learning rate of 0.001, batch size of 512, and binary cross entropy as loss function.

\begin{figure}[htp!]
\centering
\includegraphics[width=152mm]{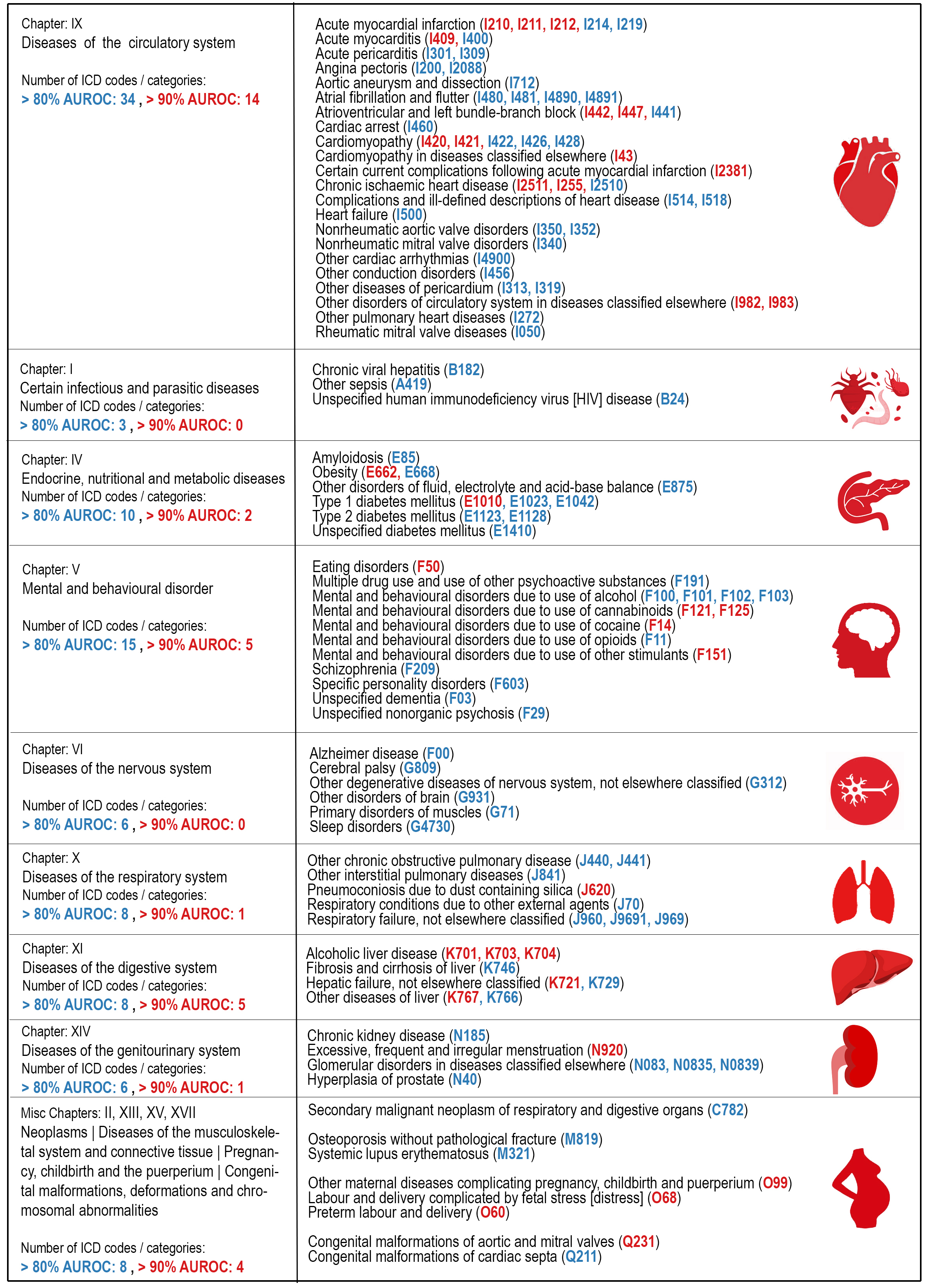}
\caption{Validated list of top performing ICD codes and categories that could be predicted from the patient's first in-hospital ECG using deep learning.}
\label{table1}
\end{figure}

\section{Results}

In our internal validation, we found 369 out of 1319 ICD codes and 170 out of 699 ICD categories to have AUROC > 80\%. Among these, 70 ICD codes and 29 ICD categories had AUROC > 90\%. However, several of these labels had low precision, therefore we restricted the list to the labels with at least 5\% AUPRC (area under precision-recall curve) or with an average precision that is at least 20 times greater than the prevalence of the condition. This yielded 151 ICD codes and 80 ICD categories with AUROC > 80\%; and 52 ICD codes and 18 ICD categories with AUROC > 90\%. Finally, we examined the replication of these lists in the external validation, and found that 128 out of 151 (84.8\%) ICD codes and 68 out of 80 (85.0\%) ICD categories were replicated to have AUROC > 80\%; and 40 out of 52 (76.9\%) ICD codes and 16 out of 18 (88.9\%) ICD categories were replicated to have AUROC > 90\%. We present this final validated list in Figure \ref{table1}, under different ICD sections.

\section{Discussion}

To the best of our knowledge, this is the first study which explores the ECG based predictability of multiple diseases over the ICD-wide diagnostic landscape. Our DL models trained and validated on population scale datasets demonstrate excellent AUROC (i.e high sensitivity \& specificity) for several diseases, however their precision (PPV) might be limited, partially owing to their low prevalence rates (89.8\% of diseases had <1\% positive class ECGs) \cite{wong2021classification}. Therefore, model predictions for such diseases might be more suitable for ‘rule out’ screening, rather than ‘rule in’ diagnostics. Population-based records enable learning from high volume healthcare data, although diagnostic labels obtained from these records may not be considered as gold-standard ground truth without proper adjudication. Also, like any other DL model, the latent ECG features used for prediction in our models may not be directly related to underlying pathology of diseases, and could be attributed to patient's comorbidities, medication usage and lifestyle factors that are naturally correlated with disease states in the population. Finally, although most adult patients get an ECG at some point during their lifetime, there is a potential for selection bias in our cohort as it is restricted to patients who had undergone at least one ECG in the 13 year period (2007-2020). Therefore, these results should be considered preliminary proof-of-concept for further investigation of specific diseases by future studies. Our next steps will focus on ensuring the generalizability across the hospitals (evaluating on ECGs from hospital sites that were not used during training) and robustness against possible model biases, such as towards certain gender or ethnic groups. That said, the current study demonstrates an exciting potential for state-of-the-art DL models trained on a ubiquitous diagnostic test (ECG) linked to routinely collected health data to transform high-throughput diagnostics for a wide range of diseases.

\section{Potential negative societal impact}

Our results are primarily based on AUROC which is commonly used in clinical models as concordance index, even though it does not account for real world costs associated with individual diseases. Moreover, given the low precision in identifying the majority of disease conditions, a positive result may trigger a cascade of invasive or non-invasive diagnostic measures that may inflict high costs on the healthcare systems. However, early diagnosis of conditions could provide opportunity to control disease progress and prevent further complications.

\paragraph{Acknowledgements}

This study is funded by Canadian Institutes of Health Research grant.

\bibliographystyle{plain}
\bibliography{neurips_2021}

\end{document}